  \documentclass{gretsi}
 \usepackage[latin1]{inputenc}   % UNIX, codage ISO 8859-1
 \usepackage[english,french]{babel}   % "babel.sty" + "french.sty"
  \usepackage{times}			% ajout times le 30 mai 2003
  \usepackage{cite} 
\usepackage{amsthm,amsmath, amsfonts, amssymb} %\usepackage{algorithm, algpseudocode}
\newtheorem{theorem}{Theorem}[section]
\newtheorem{proposition}{Proposition}[section]
\usepackage{enumitem}
\usepackage{subfig}
\usepackage{xcolor}
\usepackage{bm,graphicx, enumerate}
\usepackage{verbatim}
\usepackage{algorithm, algorithmic}
\usepackage{dsfont}

\newcommand{\modNT}[1]{#1}

\newcommand{\Graph}{\ensuremath{\set{G}}}

\newcommand{\Lap}{\ensuremath{\ma{L}}}
\newcommand{\Fou}{\ensuremath{\ma{U}}}
\newcommand{\Meas}{\ensuremath{\ma{M}}}
\newcommand{\sig}{\ensuremath{\vec{x}}}
\newcommand{\meas}{\ensuremath{\vec{y}}}
\newcommand{\nbVert}{\ensuremath{N}}
\newcommand{\nbClass}{\ensuremath{k}}
\newcommand{\Rbb}{\ensuremath{\mathbb{R}}} 
\newcommand{\Nbb}{\ensuremath{\mathbb{N}}}

\renewcommand{\leq}{\ensuremath{\leqslant}}
\renewcommand{\geq}{\ensuremath{\geqslant}}

\newcommand{\adjoint}{\ensuremath{{\intercal}}}

\newcommand{\norm}[1]{\ensuremath{\left\| #1\right\|}}

\newcommand{\ma}[1]{\ensuremath{\mathsf{#1}}}
\renewcommand{\vec}[1]{\ensuremath{\bm{#1}}}

\newcommand{\set}[1]{\ensuremath{\mathcal{#1}}}

\newcommand{\spann}{\ensuremath{{\rm span}}}

\newcommand{\ie}{\textit{i.e.}}

\newtheorem{definition}[theorem]{Definition}

\usepackage{bm}
 
\titre{\'Echantillonnage de signaux sur graphes \\via des processus dÈterminantaux}
\auteur{\coord{Nicolas}{Tremblay}{},
        \coord{Simon}{Barthelm\'e}{},
    \coord{Pierre-Olivier}{Amblard}{}}
\adresse{\affil{}{GIPSA-LAB, CNRS, Grenoble, France}}
\email{prenom.nom@gipsa-lab.fr\vspace{-0.2cm}}
\resumefrancais{Nous considÈrons l'Èchantillonnage de signaux sur graphe ‡ bande limitÈe $k$, \ie, les   combinaisons linÈaires des $k$ premiers modes de Fourier du graphe. Il existe $k$ n\oe uds du  graphe qui permettent leur reconstruction parfaite, les trouver nÈcessite cependant  une diagonalisation partielle de la matrice laplacienne, trop co˚teuse ‡ grande dimension. \modNT{
Nous proposons une nouvelle mÈthode rapide d'Èchantillonnage basÈe sur des processus dÈterminantaux qui permet la reconstruction ‡ partir d'un nombre d'Èchantillons de l'ordre de $k$.}} %Ce faisant, nous proposons un nouvel algorithme gÈnÈral d'Èchantillonnage de processus dÈterminantaux, plus rapide d'un facteur $k$ que l'algorithme de l'Ètat de l'art.}
\resumeanglais{We consider the problem of sampling $k$-bandlimited graph signals, \ie, linear combinations of the first $k$ graph Fourier modes. We know that a set of $k$ nodes embedding all $k$-bandlimited signals always exists, thereby enabling their perfect reconstruction after sampling. Unfortunately, to exhibit such a set, one needs to partially diagonalize the graph Laplacian, which becomes prohibitive at large scale. \modNT{We propose a novel strategy based on determinantal point processes that side-steps partial diagonalisation and enables reconstruction with only $O(k)$ samples.}} %While doing so, we exhibit a new general algorithm to sample determinantal process, faster than the state-of-the-art algorithm by an order $k$. }

\begin{document}
\maketitle

\noindent {\large \bf 1 \quad Introduction}
\vspace{.2cm}

%\section{Introduction}
\'Etant donnÈe une certaine classe de signaux, Èchantillonner consiste ‡ mesurer un signal un nombre de fois suffisant pour une certaine t‚che, e.g. ‡ des fins de reconstruction. Pour les signaux dÈfinis sur des graphes, une classe de rÈgularitÈ souvent utilisÈe est celle des signaux ‡ bande limitÈe $k$ (cf la dÈfini\-tion 1). Dans ce contexte, il existe deux types de mÈtho\-des d'Èchantil\-lon\-nage : i)~celles qui calculent les $k$ premiers modes de Fourier du graphe et cherchent via des heuristiques $k$ n\oe uds qui les discriminent tous~\cite{chen_sampling}, ii)~celles qui s'affranchissent de ce calcul, et qui soit cherchent un nombre de n\oe uds proche de $k$ et rÈsolvent des problËmes combinatoires qui ne passent pas ‡ l'Èchelle~\cite{anis_efficient_2016}, soit s'autorisent un nombre de n\oe uds de l'ordre de $O(k\log{k})$ et permettent de passer ‡ l'Èchelle via un Èchantil\-lon\-nage iid adaptÈ au graphe~\cite{puy_random_2016}, soit via des marches alÈatoires sur graphe spÈcifiques~\cite{tremblay_EUSIPCO2017}. 

\noindent\textbf{Contributions.} Nous proposons une mÈthode basÈe sur les processus ponctuels dÈterminantaux (PPD)~\cite{kulesza_determinantal_2012}, des processus alÈa\-toi\-res connus pour favoriser la diversitÈ des Èchantillons. \modNT{Nous adaptons un algorithme d'Èchantillonnage de PPD, et montrons comment cet algorithme peut Ítre utilisÈ} dans le cadre de l'Èchan\-til\-lon\-nage des signaux ‡ bande limitÈe sur graphe.\\

\noindent {\large \bf 2 \quad Notations et objectif}
\vspace{.2cm}

Les matrices sont en majuscule, e.g. $\ma{K}$; les vecteurs en minuscules et en gras, e.g. $\sig$; les ensembles en cursive, e.g. $\mathcal{A}$;  $\ma{K}_{\mathcal{A},\mathcal{B}}$ est la restriction de $\ma{K}$ aux lignes (resp. colonnes) indexÈes par les ÈlÈments de $\mathcal{A}$ (resp. $\mathcal{B}$); enfin : $\ma{K}_\mathcal{A}=\ma{K}_{\mathcal{A},\mathcal{A}}$. On note $\vec{\delta}_{s}$ le vecteur dont les entrÈes sont nulles sauf en $s$, et $\ma{I}_n$ la matrice identitÈ de dimension $n$. 
Soit $\Graph$ un graphe composÈ de $N$ n\oe uds interconnectÈs selon la matrice d'adjacence $\ma{W}\in\mathbb{R}^{N\times N}$ tel que $\ma{W}_{ij}= \ma{W}_{ji} \geq 0$ reprÈsente le poids associÈ au lien connectant $i$ ‡ $j$. Notons $\ma{D}$ la matrice diagonale des degrÈs: on a $\ma{D}_{ii} = \sum_{j} \ma{W}_{ij}$. Le laplacien associÈ ‡ $\Graph$ s'Ècrit $\Lap = \ma{D} - \ma{W} \in\mathbb{R}^{N\times N}$, il est semi-dÈfini positif et se diagonalise en $\Lap = \Fou\Lambda\Fou^\adjoint$, o˘ $\Fou=(\vec{u}_1|\vec{u}_2|\ldots|\vec{u}_N)\in\mathbb{R}^{N\times N}$ est la matrice de ses vecteurs propres et $\Lambda=\text{diag}(\lambda_1,\lambda_2,\ldots,\lambda_N)\in\mathbb{R}^{N\times N}$ la matrice diagonale de ses valeurs propres, rangÈes dans l'ordre croissant $0=\lambda_1\leq\lambda_2\leq\ldots\leq\lambda_N$. Par analogie au traitement du signal discret classique, $\vec{u}_i$ est considÈrÈ comme le $i$-Ëme mode de Fourier du graphe~\cite{shuman_emerging_2013}. Pour $k\in\mathbb{N}^*$, on dÈfinit $\Fou_k = (\vec{u}_1|\ldots|\vec{u}_k) \in\mathbb{R}^{N\times k}$ la concatÈnation des $k$ premiers modes de Fourier du graphe. Un signal ‡ bande limitÈe se dÈfinit alors : \\

\vspace{-0.3cm}\noindent\textbf{DÈfinition 1} (Signal ‡ bande limitÈe $k$).  
\emph{ Un signal $\sig \in \Rbb^{\nbVert}$ dÈfini sur les n\oe uds d'un graphe $\Graph$ est ‡ bande limitÈe $k\in \Nbb^*$ si $\sig \in \spann(\ma{U}_\nbClass)$, \ie, $\exists~\vec{\alpha}\in\mathbb{R}^k$ tel que $\sig = \Fou_k\vec{\alpha}$.}\\

%On s'intÈresse aux mÈthodes d'Èchantillonnage des signaux ‡ bande limitÈe $k$, qui permettent la meilleure reconstruction possible. 
\vspace{-0.3cm}En notant $m$ le nombre de mesures, Èchantillonner consiste ‡ choisir un ensemble de $m$ n\oe uds $\mathcal{A}=\{s_1,\ldots,s_m\}$. Notons $\Meas=(\vec{\delta}_{s_1}|\vec{\delta}_{s_2}|\ldots|\vec{\delta}_{s_m})^\adjoint\in\mathbb{R}^{m\times N}$ la matrice de mesure associÈe: le signal $\sig\in\mathbb{R}^{N}$ mesurÈ en $\mathcal{A}$ s'Ècrit
\begin{equation}
\label{eq:noise}
 \meas = \Meas\sig + \vec{n}\in\mathbb{R}^m,
\end{equation}
o˘ $\vec{n}\in\mathbb{R}^m$ est un bruit de mesure. 
\'Etant donnÈ que $\sig$ est supposÈ Ítre ‡ bande limitÈe $k$, on le reconstruit ‡ partir de sa mesure $\meas$ en calculant 
$
 \sig_{\text{rec}} = \Fou_k (\Meas\Fou_k)^{\dag}\meas, %= \argmin_{\vec{z}\in\spann{(\Fou_k)}}\norm{\meas - \Meas\vec{z}}^2 
$
o˘ $(\Meas\Fou_k)^{\dag}$ est le pseudo-inverse de Moore-Penrose de $\Meas\Fou_k\in\mathbb{R}^{m\times k}$. Notons $\sigma_1\leq\ldots\leq\sigma_k$ les valeurs singuliËres de $\Meas\Fou_k$. 
Le signal $\sig$ est parfaitement reconstructible (au bruit \modNT{prËs}) ‡ partir de $\meas$  si $\Fou_k^\adjoint\Meas^\adjoint\Meas\Fou_k$ est inversible, \ie, si $\sigma_1^2>0$. Dans ce cas :
\begin{align}
 \sig_{\text{rec}} &= \Fou_k (\Fou_k^\adjoint\Meas^\adjoint\Meas\Fou_k)^{-1} \Fou_k^\adjoint\Meas^\adjoint \meas\label{eq:rec}\\
 &= \sig + \Fou_k (\Fou_k^\adjoint\Meas^\adjoint\Meas\Fou_k)^{-1} \Fou_k^\adjoint\Meas^\adjoint\vec{n}.
\end{align}
Parmi tous les choix possibles de $\mathcal{A}$ (et donc de $\Meas$) qui vÈrifient $\sigma_1^2>0$, lesquels sont optimaux? Il existe plusieurs dÈfinitions d'optimalitÈ~\cite{tremblay_EUSIPCO2017}, nous nous intÈressons ici ‡ la suivante :\vspace{-0.1cm}
\begin{align}
\label{eq:optim_MV}
\displaystyle \mathcal{A}^{\text{\footnotesize{MV}}} = \arg\max_{\hspace{-0.7cm}\mathcal{A} \,\text{s.t.}\, |\mathcal{A}|=k}  \prod_{i=1}^k \sigma_i^2,\vspace{-0.6cm}
\end{align}
o˘ ``MV'' signifie ``volume maximal'': en effet, en maximisant le produit des valeurs singuliËres, on maximise le dÈterminant de $\Fou_k^\adjoint\Meas^\adjoint\Meas\Fou_k$, c'est-‡-dire le volume formÈ par les $k$ lignes ÈchantillonnÈes de $\Fou_k$. Trouver $\mathcal{A}^{\text{\footnotesize MV}}$ est NP-complet~\cite{civril_selecting_2009}. Notre objectif est de s'en approcher, c'est-‡-dire: trouver un ensemble de taille $k$ (comme $\mathcal{A}^{\text{\footnotesize MV}}$) ou proche de $k$ tel qu'on ait la meil\-leure reconstruction possible des signaux ‡ bande limitÈe $k$. \\\vspace{-.20cm}

\noindent {\large \bf 3 \quad Processus dÈterminantaux}
\vspace{.1cm}

Notons $[N]$ l'ensemble des sous-ensembles de $\{1,\ldots,N\}$ et 
$\ma{K}\in\mathbb{R}^{N\times N}$ une matrice semi-dÈfinie positive (SDP). On convient que $\text{det}(\emptyset)=1$. \\\vspace{-0.3cm}

\noindent\textbf{DÈfinition 2} ($m$-PPD)\textbf{.}  
\emph{ ConsidÈrons un processus ponctuel, \ie, un processus qui tire alÈatoirement un ensemble $\mathcal{A}\in[N]$.  Ce processus est un $m$-PPD ‡ noyau $\ma{K}$ si :\\
i)~$\mathbb{P}(\mathcal{A}) = 0$, pour tout $\mathcal{A}$ tel que $|\mathcal{A}|\neq m$. \\ 
ii)~$\mathbb{P}(\mathcal{A}) = \frac{1}{Z} \text{det}(\ma{K}_{\mathcal{A}})$, pour tout $\mathcal{A}$ tel que $|\mathcal{A}|= m$, o˘ $Z$ est la constante de normalisation.}\\\vspace{-0.3cm}

\noindent\textbf{Proposition 1.}  
\emph{Si $\ma{K}$ est un noyau de projection, \ie, $\ma{K}=\ma{X}\ma{X}^\adjoint$ avec $\ma{X}\in\mathbb{R}^{N\times d}$ et $\ma{X}^\adjoint\ma{X}=\ma{I}_d$, et si $d\geq m$, alors l'algorithme~\ref{alg:sampling_m-DPP} Èchantillonne un $m$-PPD ‡ noyau $\ma{K}$.}\\\vspace{-0.6cm}

 \begin{algorithm}[tb]
 \caption{\'Echantillonner un $m$-PPD ‡ noyau $\ma{K}$~\cite[Sec~2.4]{bardenet2016monte}}
 \label{alg:sampling_m-DPP}
\begin{algorithmic}
\label{alg:greedy}
\STATE \textbf{EntrÈe:} $\ma{K}, m$\\
$\mathcal{S} \leftarrow \emptyset$,  
dÈfinir $\vec{p}_0 = \text{diag}(\ma{K})\in\mathbb{R}^N$\\
$\vec{p} \leftarrow \vec{p}_0$\\
\textbf{for} $n=1,\ldots,m$ \textbf{do}:\\
\hspace{0.5cm}$\bm{\cdot}$ Tirer $s_n$ avec probabilitÈ $\mathbb{P}(s)=p(s)/\sum_{i}p(i)$\\
\hspace{0.5cm}$\bm{\cdot}$ $\mathcal{S} \leftarrow \mathcal{S}\cup\{s_n\}$\\
\hspace{0.5cm}$\bm{\cdot}$ Mettre ‡ jour $\vec{p}$ :
$\forall i\quad p(i) = p_0(i) - \ma{K}_{\mathcal{S},i}^\adjoint \ma{K}_{\mathcal{S}}^{-1} \ma{K}_{\mathcal{S},i}$\\
\textbf{end for}\\
\textbf{Sortie:} $\mathcal{A}\leftarrow\mathcal{S}$.
\end{algorithmic}
\vspace{-0.1cm}
\end{algorithm}
%
%
%Notons qu'il est impossible de tirer un $m$-PPD ‡ noyau $\ma{K}$ dont le rang $d$ est strictement infÈrieur ‡ $m$. En effet, supposons que $d<m$. \'Etant donnÈ que le rang d'une sous-matrice de rang $r$ est nÈcessairement infÈrieur ou Ègal ‡ $r$, le rang de $\ma{K}_{\mathcal{A}}$ est infÈrieur ou Ègal ‡ $d$, donc infÈrieur ‡ $m$. Or, $\ma{K}_{\mathcal{A}}$ est de taille $m\times m$. Le dÈterminant d'une matrice de rang dÈficient Ètant forcÈment nul, la probabilitÈ associÈe ‡ tout $\mathcal{A}$ de taille $m$ est nulle.  Dans la suite, on supposera donc qu'on a toujours  $m\leq d$. 
%
%L'algorithme~\ref{alg:sampling_m-DPP} permet d'Èchantillonner un tel processus : 

\begin{proof}
Notons \modNT{$\mathcal{S}_n$} (resp. $p_n(i)$) l'ensemble des $n$ Èchan\-tillons obtenus (resp. la valeur de $p(i)$) ‡ l'issue de l'Ètape $n$ de la boucle de l'algorithme~\ref{alg:sampling_m-DPP}. On a : $\mathcal{S}_n = \mathcal{S}_{n-1}\cup\{s_n\}$. 
En utilisant le complÈment de Schur, on a : $\forall n\in[1,m]\;,\forall i$,\vspace{-0.1cm}
 \begin{align}
 \label{eq:pn}
\text{det}\left(\ma{K}_{\mathcal{S}_{n-1}\cup\{i\}}\right) &= \left(\ma{K}_{i,i} - \ma{K}_{\mathcal{S}_{n-1},i}^\adjoint\ma{K}_{\mathcal{S}_{n-1}}^{-1}\ma{K}_{\mathcal{S}_{n-1},i}\right) \;\text{det}\left(\ma{K}_{\mathcal{S}_{n-1}}\right) \nonumber\\
& = p_{n-1}(i)\;\text{det}\left(\ma{K}_{\mathcal{S}_{n-1}}\right).\vspace{-0.3cm}
 \end{align}
 \`A partir de~\eqref{eq:pn}, et sachant que i)~$\ma{K}$ est SDP: $\forall\mathcal{S}, \,\text{det}(\ma{K}_\mathcal{S})\geq0$, ii)~$d\geq m$, on peut montrer que $p_n(i)\geq 0$ et $\sum_i{p_n}(i)\neq 0$, c'est-‡-dire qu'‡ chaque itÈration de la boucle, la probabilitÈ $\mathbb{P}(s)$ est bien dÈfinie.  
% 
%ce qui implique %, vue l'Èquation~\eqref{eq:pn}, que 
%$p_n(i)\geq 0$. De plus, $\sum_i{p_n}(i)\neq 0$. En effet, si ce n'Ètait pas le cas, cela impliquerait $\forall i, p_n(i)=0$, \ie, d'aprËs~\eqref{eq:pn}, 
%$
%\forall i, \quad\text{det}\left(\ma{K}_{\bar{\mathcal{S}}\cup\{i\}}\right)=0,
%$
%\ie, le rang de $\ma{K}$ serait infÈrieur ‡ $m$, ce qui est  contraire ‡ l'hypothËse.  \`A chaque itÈration de la boucle, la probabilitÈ $\mathbb{P}(s)$ est donc bien dÈfinie.  
La boucle Ètant rÈpÈtÈe $m$ fois, la sortie de l'algorithme, notÈe $\mathcal{A}$, est nÈcessairement de taille $m$, \modNT{en accord avec le point~i) de la DÈf.~2. Enfin, montrons que $\mathbb{P}(\mathcal{A})$ est conforme au point~ii). Par construction de $\mathcal{A}$ :\vspace{-0.0cm}
 \begin{align}
\mathbb{P}(\mathcal{A}) &= \prod_{l=1}^m \mathbb{P}(s_l|s_1, s_2,\ldots, s_{l-1})=\prod_{l=1}^m \frac{p_{l-1}(s_l)}{\sum_{i=1}^N p_{l-1}(i)}.\vspace{-0.3cm}\label{eq:prod_cond}
 \end{align}
Or, en Ècrivant~\eqref{eq:pn} pour $i=s_n$, et en itÈrant, on obtient  :
$\prod_{l=1}^m p_{l-1}(s_l)=\text{det}(\ma{K}_{\mathcal{A}})$. 
Reste ‡ montrer que le dÈnomina\-teur de~\eqref{eq:prod_cond} ne dÈpend pas des Èchantillons choisis. C'est l‡ o˘ l'hypothËse d'un noyau de projection  est essentielle. On a :\vspace{-0.2cm}
\begin{align}
 \forall l\in[1,m],  \sum_{i=1}^N p_{l-1}(i) = \sum_{i=1}^N p_{0}(i) - \sum_{i=1}^N \ma{K}_{\mathcal{S}_{l-1},i}^\adjoint\ma{K}_{\mathcal{S}_{l-1}}^{-1}\ma{K}_{\mathcal{S}_{l-1},i}\vspace{-0.3cm}\nonumber
\end{align}
On a $\sum_{i=1}^N p_{0}(i)=\text{Tr}(\ma{XX}^\adjoint)=\text{Tr}(\ma{X}^\adjoint\ma{X})=d$. De plus, soit $\ma{M}$ la matrice de mesure associÈe ‡ $\mathcal{S}_{l-1}$:\vspace{-0.2cm}
\begin{align}
 \sum_{i=1}^N &\ma{K}_{\mathcal{S}_{l-1},i}^\adjoint\ma{K}_{\mathcal{S}_{l-1}}^{-1}\ma{K}_{\mathcal{S}_{l-1},i} = \text{Tr}\left(\ma{XX}^\adjoint\ma{M}^\adjoint (\ma{MXX}^\adjoint\ma{M}^\adjoint)^{-1}\ma{MXX}^\adjoint\right)\nonumber\\
 &=\text{Tr}\left((\ma{MXX}^\adjoint\ma{M}^\adjoint)^{-1}\ma{MXX}^\adjoint\ma{XX}^\adjoint\ma{M} \right) = \text{Tr}(\ma{I}_{l-1}) = l-1\nonumber,\vspace{-0.3cm}
\end{align}
par invariance de la trace aux permutations circulaires. Ainsi :\vspace{-0.2cm}
\begin{align}
 \mathbb{P}(\mathcal{A}) = \frac{1}{Z} \text{det}(\ma{K}_{\mathcal{A}}) \text{~avec~} Z = \prod_{l=1}^m d-l+1,\vspace{-0.3cm}
\end{align}
ce qui termine la preuve.}
 \end{proof}
 
$\mathcal{A}^{\text{\footnotesize MV}}$, l'ensemble ‡ approcher, est l'ensemble le plus probable du $m$-PPD (avec $m=k$) ‡ noyau $\ma{K}_k = \Fou_k \Fou_k^\adjoint$. Si nous avons les ressources pour calculer $\Fou_k$, une premiËre stratÈgie d'Èchantillonnage est donc l'Alg.~\ref{alg:sampling_m-DPP} appliquÈ ‡ $\ma{K}_k$. Dans le cas contraire, nous proposons dans la suite un nouvel algorithme d'Èchantillonnage de $m$-PPD, de complexitÈ infÈrieure d'un facteur $m$, qui permet d'appliquer des techniques d'approximation polynomiale, Èvi\-tant ainsi toute Ètape de diagonalisation. \\% partielle du laplacien. \\

\begin{algorithm}[tb]
 \caption{\'Echantillonner un $m$-PPD, algorithme Èquivalent}
 \label{alg:sampling_m-DPP_bis}
\begin{algorithmic}
\STATE \textbf{EntrÈe:} $\ma{K}=[\vec{k}_1,\ldots, \vec{k}_N], m$\\
$\mathcal{S} \leftarrow \emptyset$, 
dÈfinir $\vec{p} = \text{diag}(\ma{K})\in\mathbb{R}^N$\\
\textbf{for} $n=1,\ldots,m$ \textbf{do}:\\
\hspace{0.5cm}$\bm{\cdot}$ Tirer $s_n$ avec probabilitÈ $\mathbb{P}(s)=p(s)/\sum_{i}p(i)$\\
\hspace{0.5cm}$\bm{\cdot}$ $\mathcal{S} \leftarrow \mathcal{S}\cup\{s_n\}$\\
\hspace{0.5cm}$\bm{\cdot}$ Calculer 
$\vec{f}_n = \vec{k}_{s_n} - \sum_{l=1}^{n-1} \vec{f}_l f_l(s_n)$\\
\hspace{0.5cm}$\bm{\cdot}$ Normaliser $\vec{f}_n \leftarrow \vec{f}_n / \sqrt{f_n(s_n)}$\\
\hspace{0.5cm}$\bm{\cdot}$ Mettre ‡ jour $\vec{p}$ : $\forall i\quad p(i) \leftarrow p(i) - f_n(i)^2$\\
\textbf{end for}\\
\textbf{Sortie:} $\mathcal{A}\leftarrow\mathcal{S}$.
\end{algorithmic}
\vspace{-0.1cm}
\end{algorithm}

\noindent {\large \bf 4 \quad Approximation via des filtres sur graphe}
\vspace{.2cm}

\noindent{\bf 4.1 \quad RÈÈcriture de l'algorithme d'Èchantillonnage}
\vspace{.1cm}
%\section{}
%Nous cherchons ‡ approcher le tirage d'un $m$-PPD ‡ noyau $\ma{K}_k$, dans les cas o˘ $N$ et/ou $k$ sont trop grands pour calculer $\Fou_k$ en un temps raisonnable. 
%\subsection{}

\noindent\textbf{Proposition 2.}  
\emph{  L'algorithme~\ref{alg:sampling_m-DPP_bis} est Èquivalent ‡ l'algorithme~\ref{alg:sampling_m-DPP} : il Èchantillonne aussi un $m$-PPD ‡ noyau de projection $\ma{K}$.}\\\vspace{-0.6cm}
\begin{proof}
ConsidÈrons $\mathcal{S}_n, \mathcal{S}_{n-1}, p_n(i)$ dÈfinis comme prÈcÈdemment et montrons que les $p_n(i)$ dans les boucles des deux algorithmes sont Ègaux. 
Dans l'Alg.~\ref{alg:sampling_m-DPP_bis} : $p_n(i) = p_{n-1}(i) - f_n(i)^2 = p_0(i) - \sum_{l=1}^{n} f_l(i)^2$ \modNT{(o˘ les $\{\vec{f}_i\}$ sont dÈfinis dans l'algorithme; notamment: $\vec{f}_1=\vec{k}_{s_1}$)}. En comparant avec  $p_n(i)$ obtenu dans l'Alg.~\ref{alg:sampling_m-DPP}, il suffit de montrer que :\vspace{-0.2cm}
 \begin{align}
 \forall n\forall i \quad \quad
  \sum_{l=1}^{n} f_l(i)^2 = \ma{K}_{\mathcal{S}_n,i}^\adjoint \ma{K}_{\mathcal{S}_n}^{-1} \ma{K}_{\mathcal{S}_n,i}.\vspace{-0.3cm}
 \end{align}
 Nous allons montrer plus gÈnÈralement que :\vspace{-0.2cm}
  \begin{align}
 \label{eq:to_proove}
\forall n \forall i,j \quad \quad
  \sum_{l=1}^{n} f_l(i)f_l(j) = \ma{K}_{\mathcal{S}_n,i}^\adjoint \ma{K}_{\mathcal{S}_n}^{-1} \ma{K}_{\mathcal{S}_n,j}.\vspace{-0.3cm}
 \end{align}
 Pour ce faire, nous proposons une rÈcurrence. \\\textit{Initialisation}. C'est vrai pour $n=1$, o˘ $\mathcal{S}_n$ est rÈduit ‡ $\{s_1\}$ :\vspace{-0.2cm}
 \begin{align}
\forall i,j\quad  \ma{K}_{\mathcal{S}_n,i}^\adjoint \ma{K}_{\mathcal{S}_n}^{-1} \ma{K}_{\mathcal{S}_n,j}= \ma{K}_{s_1,i}\ma{K}_{s_1,j} / \ma{K}_{s_1,s_1} = f_1(i) f_1(j).\nonumber\vspace{-0.3cm}
 \end{align}
\textit{HypothËse}. Supposons que~\eqref{eq:to_proove} est vraie ‡ l'Ètape $n-1$.\\%, \ie :\vspace{-0.2cm}
 %\begin{align}
% \label{eq:recbis}
%\forall i,j \quad \quad  \sum_{l=1}^{n-1} f_l(i)f_l(j) = \ma{K}_{\mathcal{S}_{n-1},i}^\adjoint \ma{K}_{\mathcal{S}_{n-1}}^{-1} %\ma{K}_{\mathcal{S}_{n-1},j}.\vspace{-0.3cm}
% \end{align}
\textit{RÈcurrence}. Montrons qu'elle est Ègalement vraie ‡ l'Ètape $n$.  En utilisant l'identitÈ de Woodbury sur $\ma{K}_{\mathcal{S}_n}^{-1}$, on montre que:\vspace{-0.2cm}
 \begin{align}
  \ma{K}_{\mathcal{S}_n,i}^\adjoint \ma{K}_{\mathcal{S}_n}^{-1} \ma{K}_{\mathcal{S}_n,j} &=
  \ma{K}_{\mathcal{S}_{n-1},i}^\adjoint \ma{K}_{\mathcal{S}_{n-1}}^{-1} \ma{K}_{\mathcal{S}_{n-1},j} + \frac{z_n(i)z_n(j)}{z_n(s_n)},\nonumber\vspace{-0.3cm}
   \end{align}
   o˘ $z_n(i)=\ma{K}_{s_{n},i}-\ma{K}_{\mathcal{S}_{n-1},s_n}^\adjoint \ma{K}_{\mathcal{S}_{n-1}}^{-1}\ma{K}_{\mathcal{S}_{n-1},i}$. En remplaÁant $\ma{K}_{\mathcal{S}_{n-1},i}^\adjoint \ma{K}_{\mathcal{S}_{n-1}}^{-1} \ma{K}_{\mathcal{S}_{n-1},j}$ par $\sum_{l=1}^{n-1} f_l(i)f_l(j)$ gr‚ce ‡ l'hypothËse, il nous reste ‡ montrer que   :\vspace{-0.2cm}
%    \begin{align}
%   \ma{K}_{\mathcal{S},i}^\adjoint \ma{K}_{\mathcal{S},\mathcal{S}}^{-1} \ma{K}_{\mathcal{S},j} &= \sum_{l=1}^{n-1} f_l(i)f_l(j) + \frac{z_n(i)z_n(j)}{z_n(s_n)}. 
% \end{align}
% Il nous reste donc ‡ montrer que :
\begin{align}
\label{eq:reste_a_montrer}
\forall i,j \quad \quad f_{n}(i)f_{n}(j) = \frac{z_n(i)z_n(j)}{z_n(s_n)}.\vspace{-0.3cm}
\end{align}
Or, par construction dans l'Algorithme~\ref{alg:sampling_m-DPP_bis}, $f_n(i)$ s'Ècrit :\vspace{-0.2cm}
 \begin{align}
 \label{eq:f}
\forall i\quad\quad f_{n}(i) = \frac{\ma{K}_{s_{n},i} - \sum_{l=1}^{n-1}f_l(i)f_l(s_{n})}{\sqrt{\ma{K}_{s_{n}} - \sum_{l=1}^{n-1}f_l(s_{n})^2}}.\vspace{-0.3cm}
\end{align}
En utilisant une seconde fois l'hypothËse~\eqref{eq:rec}, on montre que :\vspace{-0.2cm}
%
%
%on a :
% \begin{align}
%\forall i\quad\quad f_{n}(i) = \frac{\ma{K}_{s_{n},i} - \ma{K}_{\bar{\mathcal{S}},s_n}^\adjoint \ma{K}_{\bar{\mathcal{S}},%\bar{\mathcal{S}}}^{-1}\ma{K}_{\bar{\mathcal{S}},i}}{\sqrt{\ma{K}_{s_{n},s_{n}} - \ma{K}_{\bar{\mathcal{S}},s_n}^\adjoint \ma{K}_{\bar{\mathcal{S}},%\bar{\mathcal{S}}}^{-1}\ma{K}_{\bar{\mathcal{S}},s_n}}}.
%\end{align}
%Si bien que :
\begin{align}
\forall i\quad\quad 
 f_{n}(i) = \frac{z_n(i)}{\sqrt{z_n(s_n)}},\vspace{-0.3cm}
\end{align}
ce qui prouve~\eqref{eq:reste_a_montrer} et termine la preuve.
\end{proof}
L'algorithme~\ref{alg:sampling_m-DPP_bis} est un algorithme gÈnÈral pour Èchantillonner un $m$-PPD ‡ noyau de projection. Sa complexitÈ est en $O(Nm^2)$, alors que la complexitÈ de l'algorithme~\ref{alg:sampling_m-DPP} est en $O(Nm^3)$. Notons que des idÈes similaires pour rÈduire la complexitÈ d'un facteur $m$ existent dans la littÈrature (par exemple dans~\cite{lavancier_determinantal_2015}) mais sous des formes un peu cachÈes, et, ‡ notre connaissance, n'ont jamais ÈtÈ vraiment explicitÈes dans le cas discret. \\

\noindent{\bf 4.2 \quad Approcher l'Èchantillonnage d'un PPD}
\vspace{0.2cm}

%\subsection{Approcher l'Èchantillonnage d'un PPD}
\label{subsec:approx}
Nous allons voir que la maniËre dont l'algorithme~\ref{alg:sampling_m-DPP_bis} fait appel ‡ $\ma{K}$ permet de tirer profit d'approximations polynomiales usuellement utilisÈes lors d'opÈ\-ra\-tions de filtrage sur graphe. Rappelons que nous considÈrons le $m$-PPD associÈ au noyau $\ma{K}_k$. Or :
$
 \ma{K}_k = \Fou h_k(\ma{\Lambda})\Fou^\adjoint,
$
o˘ $h_k(\ma{\Lambda}) = \text{diag}(h_k(\lambda_1),\ldots,h_k(\lambda_N))$ et $h_k(\lambda)$ est tel que $h_k(\lambda)=1$ si $\lambda\leq\lambda_k$ et $h_k(\lambda)=0$ sinon. %:
%\begin{align}
% h_k(\lambda)&=1 ~~~\mathrm{si} ~\lambda\leq\lambda_k\\\nonumber
% h_k(\lambda)&=0 ~~~\mathrm{sinon.}
%\end{align}
En traitement du signal sur graphes, $\ma{K}_k$ est un filtre passe-bas idÈal de frÈquence de coupure $\lambda_k$. 
\begin{algorithm}[tb]
 \caption{\'Echantillonnage approchÈ d'un $m$-PPD ‡ noyau $\ma{K}=\modNT{h_k}(\Lap)$}
 \label{alg:sampling_approx}
\begin{algorithmic}
\STATE \hspace{-0.2cm}\textbf{EntrÈe:} $\Lap$, $\modNT{h_k}(\lambda)$, $r$, $m$\\
\hspace{-0.2cm}Calculer $\lambda_N$, la plus grande valeur propre de $\Lap$\\
\hspace{-0.2cm}Calculer le polynÙme $\modNT{\tilde{h}_k}$ de degrÈ $r$ approchant $\modNT{h_k}$ sur $[0,\lambda_N]$\\
\hspace{-0.2cm}Estimer $\vec{p} =\text{diag}(\modNT{\tilde{h}_k}(\Lap))\in\mathbb{R}^N$ comme vu dans la Sec.~4.2\\
\hspace{-0.2cm}\textbf{for} $n=1,\ldots,m$ \textbf{do}:\\
\hspace{0.5cm} $\bm{\cdot}$ Tirer $s_n \leftarrow \text{argmax}(\vec{p})$\\
\hspace{0.5cm} $\bm{\cdot}$ Calculer
$\vec{f}_n = \modNT{\tilde{h}_k}(\Lap) \vec{\delta}_{s_n} - \sum_{l=1}^{n-1} \vec{f}_l f_l(s_n)$\\
\hspace{0.5cm} $\bm{\cdot}$ Normaliser $\vec{f}_n \leftarrow \vec{f}_n / \sqrt{f_n(s_n)}$\\
\hspace{0.5cm} $\bm{\cdot}$ Mettre ‡ jour $p(i) \leftarrow p(i) - f_n(i)^2$\\
\hspace{-0.2cm}\textbf{end for}\\
\hspace{-0.2cm}\textbf{Output:} $\mathcal{A} = \{s_1,\ldots,s_m\}$
\end{algorithmic}
\vspace{-0.1cm}
\end{algorithm}

\noindent\textbf{Approximation polynomiale.}~\modNT{\cite{shuman_emerging_2013}} %ConsidÈrons un filtre quelconque $h(\lambda)$ dÈfini dans l'espace de Fourier du graphe, et Ècrivons le filtre associÈ dans l'espace des n\oe uds: $\ma{H} = \Fou h(\ma{\Lambda})\Fou^\adjoint$.
ConsidÈrons le polynÙme de Tchebychev $\tilde{h}_k$ de degrÈ $r$ qui approche au mieux $h_k$:\vspace{-0.2cm} $$\forall\lambda\in[0,\lambda_N]\quad\tilde{h}_k(\lambda) = \sum_{l=1}^r \alpha_l \lambda^l \simeq h_k(\lambda).$$ \vspace{-0.3cm}
\begin{equation}
\text{On a : } \ma{K}_k \simeq \Fou \tilde{h}_k(\ma{\Lambda})\Fou^\adjoint = \sum_{l=1}^r \alpha_l \Fou\Lambda^l\Fou^\adjoint =  \sum_{l=1}^r \alpha_l \Lap^l  = \tilde{h}_k(\Lap).\nonumber\vspace{-0.1cm}
\end{equation}

\noindent\textbf{Filtrage rapide sur graphe.}
On ne calcule jamais explicitement $\tilde{h}_k(\Lap)$ qui est en gÈnÈral dense de taille $N\times N$. En revanche, Ètant donnÈ un signal $\sig$ dÈfini sur le graphe, le signal filtrÈ par $h_k$, $\ma{K}_k\sig$, est approchÈ par $\tilde{h}_k(\Lap)\sig = \sum_{l=1}^r \alpha_l \Lap^l \sig$, qui se calcule  via $r$ multiplications matrice-vecteur %(on calcule d'abord $\Lap\sig$, puis $\Lap(\Lap\sig)$, etc.) 
si bien que le nombre d'opÈrations nÈcessaires pour filtrer un signal est de l'ordre de $r|E|$ o˘ $|E|$ est le nombre de liens du graphe. 

\noindent\textbf{Estimation de la diagonale de $\ma{K}_k$.} 
%Notons $\ma{H}_{\frac{1}{2}}=\Fou\sqrt{h(\Lambda)}\Fou^\adjoint$. Soit $\tilde{h}_{\frac{1}{2}}$ l'approximation polynomiale de $\sqrt{h}$. On a : $\ma{H}_{\frac{1}{2}} \simeq \tilde{h}_{\frac{1}{2}}(\Lap)$. 
%\begin{equation}
% \ma{H}_{\frac{1}{2}} \simeq  \Fou \tilde{h}_{\frac{1}{2}}(\Lambda)\Fou^\adjoint = \sum_{l=1}^r \beta_l\Lap^l = %\tilde{h}_{\frac{1}{2}}(\Lap).
%\end{equation}
Soit $\ma{R}\in\mathbb{R}^{N\times n}$ une matrice contenant $n$ signaux alÈatoires gaussiens de moyenne nulle et de variance $1/n$. \modNT{Notons que:\vspace{-0.1cm}
 \begin{align}
\mathbb{E}&\left(\norm{\vec{\delta}_i^\adjoint\tilde{h}_{k}(\Lap)\ma{R}}^2\right) = \vec{\delta}_i^\adjoint\tilde{h}_{k}(\Lap)\mathbb{E}(\ma{R}\ma{R}^\adjoint)\tilde{h}_{k}(\Lap)\vec{\delta}_i \\
&= \vec{\delta}_i^\adjoint(\tilde{h}_{k}(\Lap))^2\vec{\delta}_i\simeq \vec{\delta}_i^\adjoint\ma{K}_k^2\vec{\delta}_i=\ma{K}_k(i,i)\vspace{-0.3cm}
 \end{align}}
La $i^\text{\footnotesize Ëme}$ valeur de la diagonale de $\ma{K}_k$, $p_0(i)$,  est donc approchÈe par la norme de la $i^\text{\footnotesize Ëme}$ ligne de $\tilde{h}_k(\Lap)\ma{R}$, \ie~: \vspace{-0.2cm}
\begin{align}
\label{eq:est_diago}
p_0(i)\simeq\norm{\vec{\delta}_i^\adjoint\tilde{h}_{k}(\Lap)\ma{R}}^2,\vspace{-0.3cm}
\end{align}
et, via le lemme de Johnson-Linden\-strauss, on montre qu'un nombre de signaux alÈatoires $n=O(\log{N})$ est suffisant pour avoir une estimation convenable avec haute probabilitÈ~\cite{tremblay_ICASSP2016}.

\begin{figure*}
 \centering
a)\hspace{-0.4cm}\includegraphics[width=0.53\columnwidth]{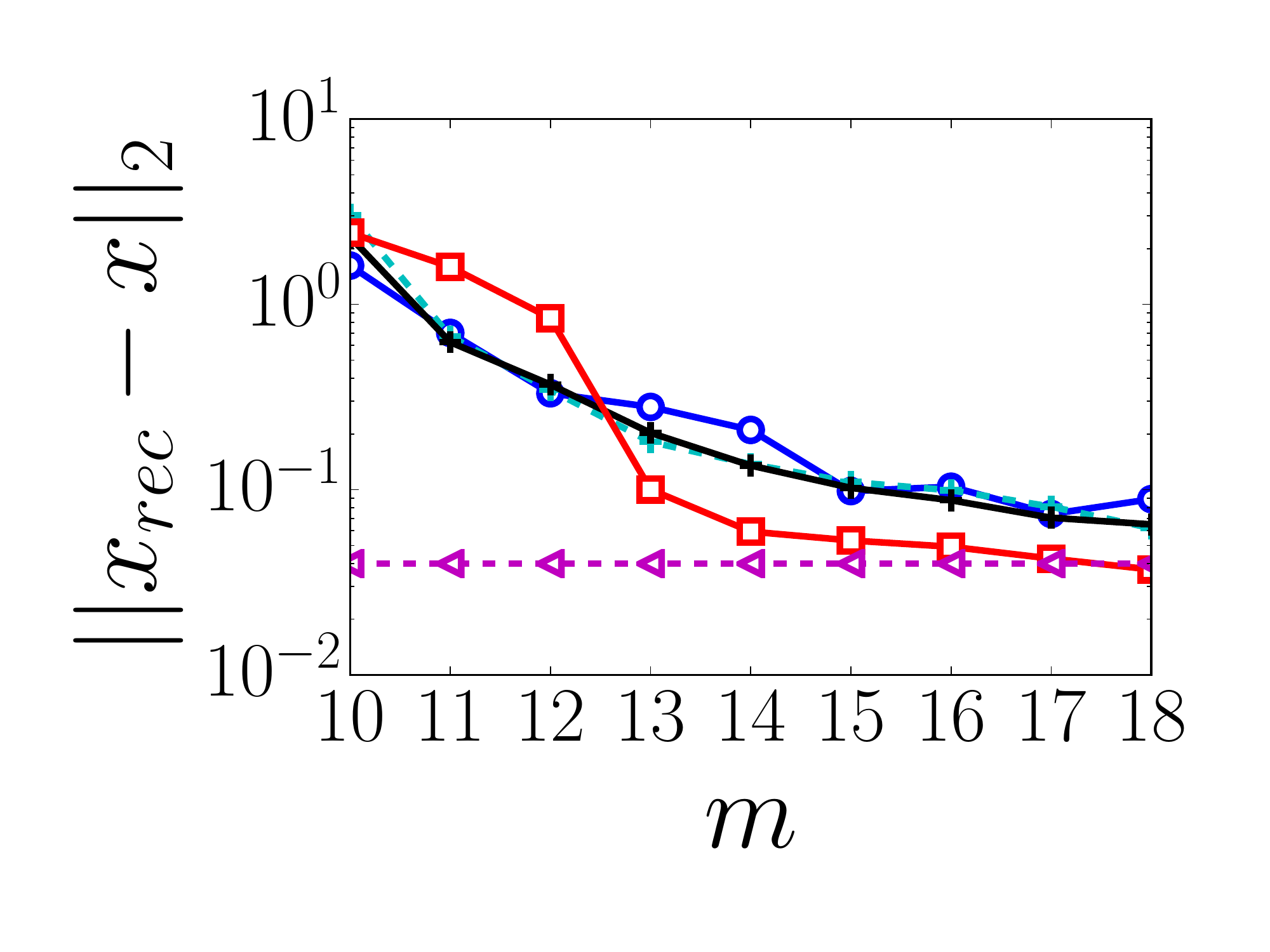}  \hspace{-0.0cm}
b)\hspace{-0.4cm}\includegraphics[width=0.53\columnwidth]{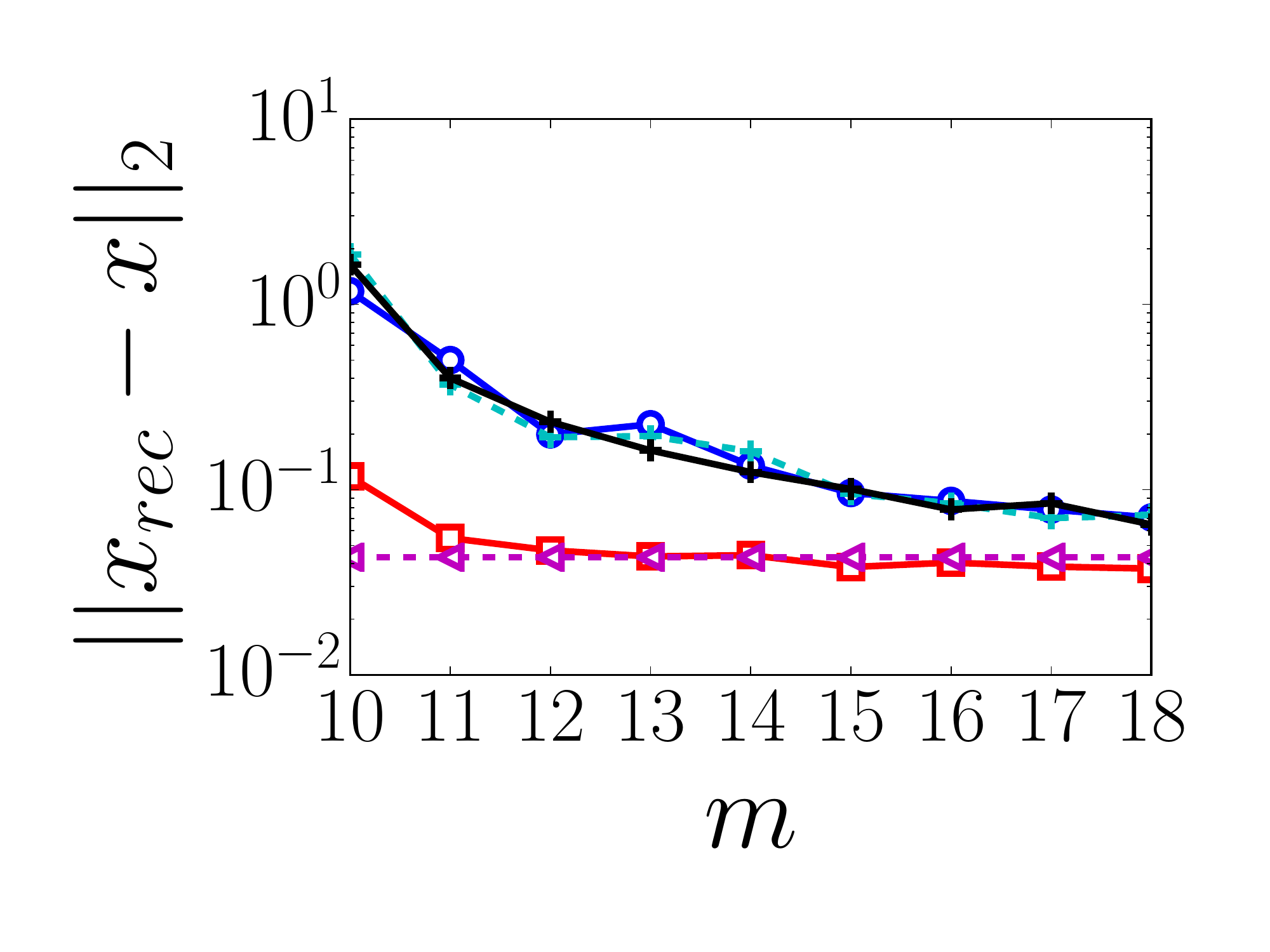}  \hspace{-0.0cm}
c)\hspace{-0.4cm}\includegraphics[width=0.8\columnwidth]{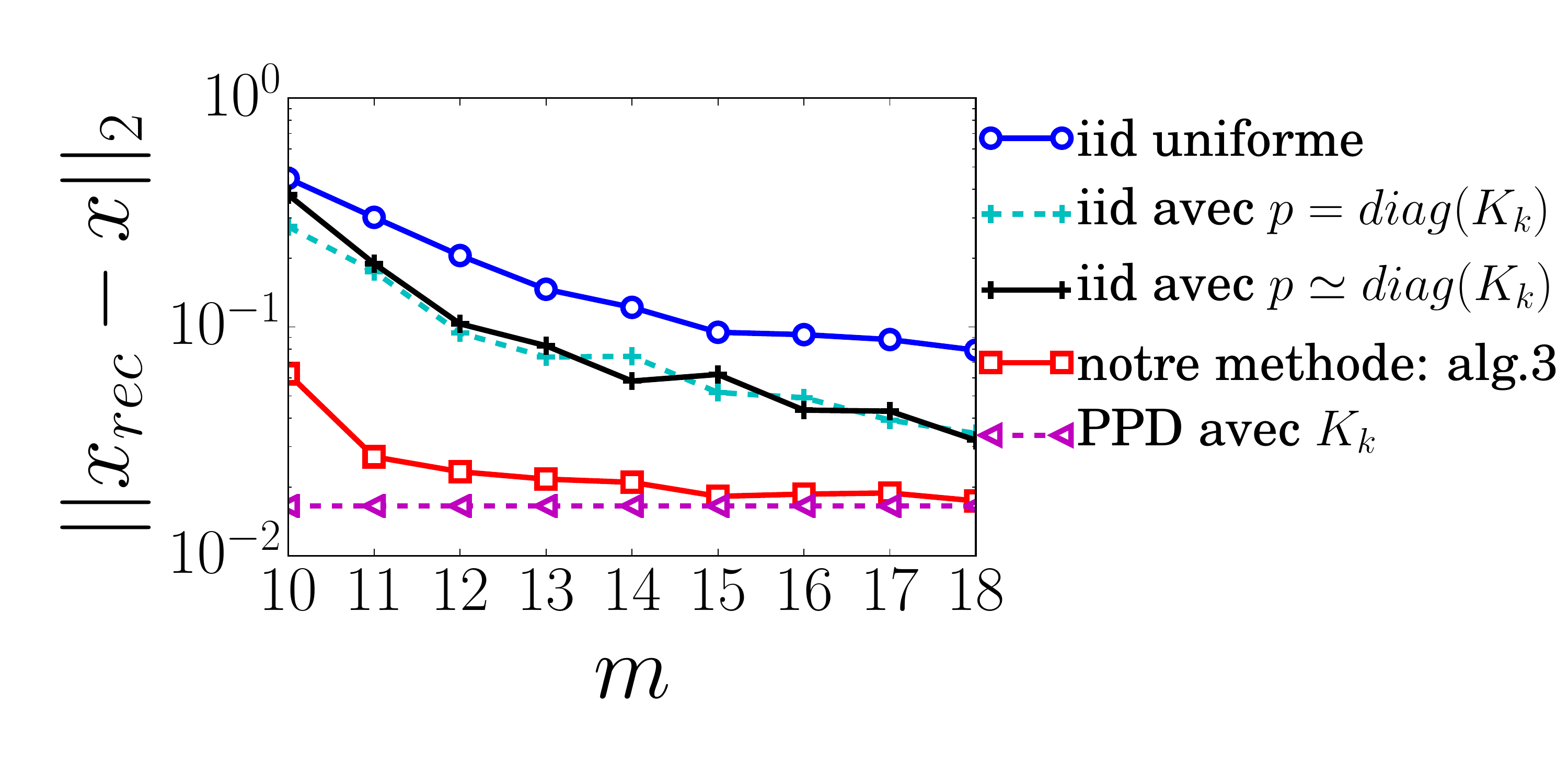}\vspace{-0.2cm} %\\
%~~~~~~~~~~~~~a) ~~~~~~~~~~~~~~~~~~~~~~~~~~~~~~~~~~ b) ~~~~~~~~~~~~~~~~~~~~~~~~~~~~~~~~~ c) ~~~~~~~~~~~~~~~~~~~~~~~~~~~~~~~d) ~~~~~~~~~~~~~~~~~~~~~~
\caption{\small{Performance mÈdiane obtenue en reconstruisant 100 signaux ‡ bande limitÈe 10 sur a)~le graphe du Minnesota, b)~le graphe ``Bunny'', c)~une rÈalisation d'un MSB. Les mÈthodes en pointillÈs doivent calculer $\Fou_k$ pour Èchantillonner, les autres non. La mÈthode via le PPD ‡ noyau $\ma{K}_k$ ne peut Èchantillonner que $m=10$ n\oe uds : la droite horizontale associÈe est artificielle et sert uniquement ‡ la comparaison. } }
\label{fig:experiments}
\end{figure*}

\noindent\textbf{Algorithme d'approximation.} On souhaite approcher l'algorithme~\ref{alg:sampling_m-DPP_bis} avec entrÈe $\ma{K}_k$ sans avoir ‡ calculer $\Fou_k$. Pour ce faire, on suit l'algorithme~\ref{alg:sampling_approx} en lui donnant comme entrÈe\footnote{pour avoir $h_k(\lambda)$, il faut $\lambda_k$, que nous estimons en suivant~\cite[Sec. 4.2]{puy_random_2016}.} $\Lap$, 
$h_k(\lambda)$, $r=50$ et $m$ le nombre d'Èchantillons souhaitÈ. Au lieu d'accÈder directement ‡ la diagonale d'un noyau $\ma{K}$ connu, on l'estime ‡ l'aide de l'Èquation~\eqref{eq:est_diago}. Puis, au lieu d'ac\-cÈ\-der aux colonnes $\vec{k}_s$ directement, on les estime via le filtrage rapide associÈ: $\vec{k}_s \simeq \tilde{h}_k(\Lap)\vec{\delta}_s$. 
Au vu des erreurs d'approximation successives, l'Ècart entre les $\{\vec{f}_l\}$ calculÈs au cours des algorithmes~\ref{alg:sampling_m-DPP_bis} et~\ref{alg:sampling_approx} ne cesse de s'accroÓtre au fur et ‡ mesure de la boucle. Pour stabiliser l'algorithme~\ref{alg:sampling_approx}~: i)~le nouvel Èchantillon est celui qui maximise $\vec{p}$ \modNT{(mise ‡ part l'estimation de la diagonale de $\ma{K}$, l'algorithme n'est donc plus alÈatoire)}; ii)~et on s'autorise $m\geq k$ (ce qui n'a pas de sens pour l'algorithme~\ref{alg:sampling_m-DPP_bis} appliquÈ ‡ $\ma{K}_k$ qui est de rang $k$). Aussi, en pratique, on fait quelques modifications mineures : i)~aprËs la mise ‡ jour de $\vec{p}$, on force $p(s_i) = 0$ pour tous les Èchantillons $s_i$ dÈj‡ choisis; ii)~lors de la normalisation de $\vec{f}_n$, si $\vec{f}_n(s_n)\leq 0$ ‡ cause de l'approximation, alors on normalise $\vec{f}_n$ par $\sqrt{\norm{\vec{f}_n}/N}$. \modNT{Ces choix sont cependant arbitraires : une comparaison rigoureuse des diffÈrentes possibilitÈs de stabilisation fera l'objet de travaux futurs.}\\

\newpage
\noindent {\large \bf 5 \quad Simulations}
%\vspace{.2cm}

%\section{Simulations}

\noindent\textbf{Trois graphes}. Nous considÈrons le graphe routier du Minnesota ($N=2642$), le graphe correspondant au maillage 3D d'un objet (graphe ``Bunny'' avec $N=2503$), et une rÈali\-sa\-tion d'un graphe alÈatoire avec communautÈs issue du modËle stochastique par blocs (MSB) avec $N=1000$, 10 communautÈs de mÍme taille et de paramËtre de structure $\epsilon = \epsilon_c/4$, o˘ $\epsilon_c$ correspond ‡ la limite o˘ les structures en communautÈs ne sont plus dÈtectables; et plus $\epsilon$ est petit, plus la structure en communautÈs est forte. Voir par exemple\cite[sec. 4]{tremblay_EUSIPCO2017} pour des dÈtails sur $\epsilon$;~\cite[Fig. 1]{puy_random_2016} pour des illustrations des deux premiers graphes. 

\noindent\textbf{Pour gÈnÈrer les signaux ‡ bande limitÈe.} Dans tous ces ex\-emples, nous choisissons $k=10$. Chaque signal $\sig$ ‡ bande limitÈe $k$ est gÈnÈrÈ en calculant $\Fou_k$, puis en calculant $\sig = \Fou_k\vec{\alpha}$ o˘ $\vec{\alpha}\in\mathbb{R}^k$ est tirÈ alÈatoirement selon une gaussienne $\mathcal{G}(0,1)$, puis en normalisant $\sig$ pour qu'il soit de norme 1. 

\noindent\textbf{Les diffÈrentes mÈthodes d'Èchantillonnage comparÈes.} On compare 5 mÈthodes d'Èchantillonnage : i)~l'Èchantillonnage iid uniforme sans remise, ii)~l'Èchantillonnage iid selon $\vec{p} = \text{diag}(\ma{K}_k)$ sans remise (comme dans~\cite{puy_random_2016}), 
iii)~la mÍme stratÈgie mais avec une distribution $\vec{p} \simeq \text{diag}(\ma{K}_k)$ estimÈe via \eqref{eq:est_diago}, iv)~l'Èchantillonnage dÈpendant en suivant l'algorithme~\ref{alg:sampling_approx}, et v)~l'Èchantillonnage selon le PPD idÈal, c'est-‡-dire en suivant l'algorithme~\ref{alg:sampling_m-DPP} avec la lÈgËre modification qu'‡ chaque itÈration de la boucle on tire l'Èchantillon qui maximise la probabilitÈ (cela donne de meilleures performances de reconstruction dans les cas testÈs). Les quatre premiËres mÈthodes peuvent Èchantil\-lon\-ner un nombre quelconque de n\oe uds $m$; la derniËre Èchantil\-lon\-ne nÈcessairement $k$ n\oe uds. 

\noindent\textbf{Le bruit de mesure} $\vec{n}\in\mathbb{R}^m$ est tirÈ selon une Gaussienne de moyenne nulle et d'Ècart-type $\sigma$, et est ajoutÈ ‡ la mesure du signal, comme dans~\eqref{eq:noise}. On fixe $\sigma=10^{-3}$. 

\noindent\textbf{Reconstruction des signaux ÈchantillonnÈs.} Quelle que soit la mÈthode choisie, on reconstruit les signaux avec~\eqref{eq:rec}, c'est-‡-dire avec la connaissance de $\Fou_k$. Il existe des moyens de reconstruction qui ne nÈcessitent pas $\Fou_k$ (voir~\cite{puy_random_2016}); mais nous souhaitons ici comparer exclusivement l'Èchantillonnage. 

\noindent\textbf{Discussion des rÈsultats.} La Fig.~\ref{fig:experiments} montre que notre mÈthode approche la performance du PPD ‡ noyau $\ma{K}_k$ plus rapidement que les autres, dËs $m=O(k)$. Pour le graphe du Minnesota, ‡ petit $m$, notre performance est moindre. Ceci dit, nous retrouvons une meilleure performance pour d'autres choix de $k$. De nombreuses expÈriences ont ÈtÈ menÈes sur d'autres rÈalisations du MSB, avec d'autres valeurs de $k$ \modNT{(diffÈ\-rentes de $k=10$ choisie ici)}, $\epsilon$ et $\sigma$, et des communautÈs de taille hÈtÈrogËne: nous trouvons toujours des comportements similaires.  \\\vspace{-0.25cm}

\noindent {\large \bf 6 \quad Conclusion}
%\vspace{.1cm}

%\section{}
\noindent\modNT{Nous proposons l'algorithme~\ref{alg:sampling_m-DPP_bis}, une rÈÈcriture moins co˚teuse de l'algorithme~\ref{alg:sampling_m-DPP} d'Èchantillonnage de $m$-DPP ‡ noyau de projection. Dans les cas o˘ le noyau est de la forme $\ma{K}=h_k(\Lap)$ o˘ $\Lap$ est une matrice diagonalisable (de prÈfÈrence parcimonieuse), nous proposons l'algorithme~\ref{alg:sampling_approx} qui,  via des approximations polynomiales, permet d'approcher l'Èchantillonnage du $m$-PPD associÈ ‡ $\ma{K}$ sans jamais calculer explicitement $\ma{K}$, potentiellement dense, et en Èvitant tout calcul de diagonalisation de $\Lap$. Nous illustrons avec succËs l'intÈrÍt de cet algorithme pour l'Èchantil\-lon\-nage de signaux sur graphe ‡  bande limitÈe.} \\

\vspace{-0.3cm}
\noindent{\bf Remerciements.} Ce travail a ÈtÈ en partie soutenu par le LabEx PERSYVAL-Lab ({\small{ANR-11-LABX-0025-01}}), l'ANR GenGP ({\small{ANR-16-CE23-0008}}), le LIA CNRS/Melbourne Univ Geodesic.
\vspace{-0.7cm}

{
\footnotesize{
}}

\end{document}